\documentclass[aps,preprint,showpacs,showkeys]{revtex4}%
\usepackage{amsfonts}
\usepackage{subfigure}
\usepackage{amsmath}
\usepackage{amssymb}
\usepackage{graphicx}
\usepackage{epsfig}%

\providecommand{\U}[1]{\protect\rule{.1in}{.1in}}

\begin{document}
\title[ ]{\textbf{Decoherence effects on multiplayer cooperative quantum games}}
\author{Salman Khan}
\email {sksafi@phys.qau.edu.pk}, ,
\author{M. Ramzan}
\author{M. Khalid Khan}
\affiliation{Department of Physics, Quaid-i-Azam University, Islamabad 45320, Pakistan.}
\keywords {Multiplayer cooperative quantum games; payoffs, Nash equilibria.}

\begin{abstract}
We study the behavior of cooperative multiplayer quantum games [35,36] in
the presence of decoherence using different quantum channels such as
amplitude damping, depolarizing and phase damping. It is seen that the
outcomes of the games for the two damping channels with maximum values of
decoherence reduce to same value. However, in comparison to phase damping
channel, the payoffs of cooperators are strongly damped under the influence\
amplitude damping channel for\ the lower values of decoherence parameter. In
the case of depolarizing channel, the game is a no-payoff game irrespective
of the degree of entanglement in the initial state for the larger values of
decoherence parameter. The decoherence gets the cooperators worse off .

PACS:03.65.Ta; 03.65.-w; 03.67.Lx\newline
Keywords: Multiplayer cooperative quantum games; payoffs, Nash equilibria.
\end{abstract}
\maketitle

\section{Introduction}

Game theory provides a mathematical background for evaluating behavior of
competing agents. Emerged from the work of von Neumann \cite{Von},
theoreticians in various disciplines such as economics, biology, medical
sciences, social sciences and physics utilize its concepts to maneuver
competing situations.\cite{Broom1, Broom2, Hofbauer, Piotrowski, Baaquie}.
Although technically difficult, quantum theory is conceptually very rich and
quantum game theorists use it to study the behavior of classical games in
this realm for more than a decade [$7$-$16$]. The quantum extension of
classical games began from the seminal work of Meyer \cite{Meyer}. It is
shown that the quantum mechanical treatment of classical games produces
results that cannot be achieved in the classical formalism. Quantum
strategies and quantum entanglement lead quantum players to harness the
outcome of the game in their favour.

Quantum mechanically competing agents communicate with each other through
quantum channels. Information can be encoded in qubits, qutrits or qudits.
These sources of information while passing through the channels interact
with the many degrees of freedom of the environment thereby creating
entangled state with it. This leads to the distortion of system space and
results in the loss of encoded information which is not inevitable \cite%
{Zurek}. The distortion of the system space through the interaction with
environment is called decoherence. Quantum error correction \cite{Steane}
and entanglement purification \cite{Deutsch}\ are the two methods developed
to handle the problem of decoherence. Quantum games in the presence of
decoherence have been studied by a number of authors \cite{Flitney3, Salman,
Salman1, Gawron, Gawron2, Chen, Xia} and many more. It is seen that the
effect of decoherence on the payoff functions of players is different for
different games setup. For example, in some cases it gets worse off the
players while in other cases it makes better off one player over the other.%
\cite{Salman, Xia}.

In the field of quantum games most work in the beginning was done in
studying two person games. Benjamin and Hayden \cite{Benjamin2} were the
first to study multiplayer games. Few of many others who contributed to the
study of multiplayer games are given in \cite{Kay, Du, Han, Iqbal, Abbott,
Wang, Hollenberg}.

In this paper we investigate the effect of decoherence and entanglement on
cooperative three and four players quantum game under the action of
amplitude damping, depolarizing and phase damping channels. The amount of
decoherence in the case of each channel is parameterized by the decoherence
parameter $p$ which has values from the range $0$ to $1$. The lower and
upper limits of $p$ correspond to undecohered and fully decohered cases
respectively.$\,$

\section{Three-players cooperative game}

A classical three persons symmetric cooperative game consists of three
players $A$, $B$, $C$ with strategy set $U_{n}$, $\{n=A$, $B$, $C\}$ for
each player and three real valued payoff functions $P_{A}$, $P_{B}$, $P_{C}$%
, one each corresponds to a player. The strategy set of each player consists
of two strategies denoted by $0$ and $1$. Players $A$ and $B$ are said to be
cooperators if they choose the same strategy, different from $C$ in a play
of the game. No one wins if they all choose the same strategy and the loser
is one who chooses strategy different from the other two players. In each
play of the game, the loser pay a fixed amount to the other two winners that
is equally divided between them. Hence the game in its classical form is a
zero sum game.

The quantum version of the game consists of three qubits, one for each
player, that is, the game space is an eight dimensional Hilbert space. The
strategy set of each player consists of two strategies $I$ and \thinspace $%
\sigma _{x}$, where $I$ is the single qubit identity operator and $\sigma
_{x}$ is the Pauli spin flip operator. The game starts from an initial three
qubits entangled state, prepared by an arbiter. The initial state is sent to
each player. The players execute their strategies on their own qubit and the
final state is returned to the arbiter. The arbiter performs measurement in
the computational basis and the corresponding payoffs of the players are
declared.

The game was initially quantized in two different ways of using the
strategies $I$ and \thinspace $\sigma _{x}$. In Ref. \cite{Iqbal2} the
classical probability method, in which each player has the option to use $I$
with certain probability $x$ and $\sigma _{x}$ with probability $1-x$, has
been used. Whereas in Ref. \cite{Ying} the quantum superposed operator
method has been adopted in quantizing the three players game. In this
method, each player has the option to execute his strategy as a linear
combination of the two allowed strategies in the form $U_{i}=\sqrt{x}I+\sqrt{%
1-x}\sigma _{x}$. When this is applied to state $|0\rangle $, it gives $%
\sqrt{x}|0\rangle +\sqrt{1-x}|1\rangle $. This means that the player on
measurement gets $0$ with probability $x$ and $1$ with probability $1-x$.
However for four players cooperative game, both methods are used in Ref.
\cite{Ying} to quantize the game. It is shown that the both methods produce
the same outcome. Keeping this in mind, we proceed to incorporate
decoherence effects in the quantum superposed formalism both for three and
four players games.

\subsection{Quantum channels}

A quantum channel transfers information from one place (input) to another
place (output). In the course of transformation, the source of information
may interact with the many degrees of freedom of the channel and leads to
the information damage. The effect of quantum channels on the state of a
system is a completely positive trace preserving map that is described in
terms of Kraus operators \cite{Nielson}.
\begin{equation}
\rho _{f}=\sum\limits_{k}E_{k}\rho _{i}E_{k}^{\dagger },  \label{1}
\end{equation}%
where $\rho _{i}=|\psi \rangle \langle \psi |$ is the initial density matrix
of the system with $|\psi \rangle $ being the initial state. The Kraus
operators $E_{k}$ satisfy the following completeness relation%
\begin{equation}
\sum\limits_{k}E_{k}^{\dagger }E_{k}=I.  \label{2}
\end{equation}%
\begin{table*}[htb]%
\caption{A single qubit Kraus operators for amplitude damping channel, phase
damping channel and depolarizing channel. \label{table:1}}%
\begin{tabular}{|c|c|}
\hline
Amplitude damping & $E_{o}=\left(
\begin{array}{cc}
1 & 0 \\
0 & \sqrt{1-p}%
\end{array}%
\right) ,\qquad E_{1}=\left(
\begin{array}{cc}
0 & \sqrt{p} \\
0 & 0%
\end{array}%
\right) $ \\ \hline
Phase damping & $E_{o}=\left(
\begin{array}{cc}
1 & 0 \\
0 & \sqrt{1-p}%
\end{array}%
\right) ,\qquad E_{1}=\left(
\begin{array}{cc}
0 & 0 \\
0 & \sqrt{p}%
\end{array}%
\right) $ \\ \hline
Depolarizing & $%
\begin{array}{c}
E_{o}=\sqrt{1-p}\left(
\begin{array}{cc}
1 & 0 \\
0 & 1%
\end{array}%
\right) ,\qquad E_{1}=\sqrt{p/3}\left(
\begin{array}{cc}
0 & 1 \\
1 & 0%
\end{array}%
\right) , \\
E_{2}=\sqrt{p/3}\left(
\begin{array}{cc}
0 & -i \\
i & 0%
\end{array}%
\right) ,\qquad E_{3}=\sqrt{p/3}\left(
\begin{array}{cc}
1 & 0 \\
0 & -1%
\end{array}%
\right)%
\end{array}%
$ \\ \hline
\end{tabular}

\end{table*}%
The single qubit Kraus operators for different channels used in this paper
are given in Table 1. The Kraus operators for three-players and four-players
are of dimensions $2^{3}$ and $2^{4}$ respectively. These Kraus operators
are constructed by taking the tensor product of all possible combinations of
single qubit Kraus operators in the following way%
\begin{equation}
E_{k}=\bigotimes\limits_{i}E_{i}
\end{equation}%
where $E_{i}$ represent the Kraus operators of a single qubit for a given
channel and the index $i$ stands for the number of single qubit Kraus
operators for that particular channel.

For the three players game, we consider the initial state to be $|\psi
\rangle =\cos \theta /2|000\rangle +\sin \theta /2|111\rangle $, where $%
\theta \in \lbrack 0,\frac{\pi }{2}]$ is a measure of entanglement \cite%
{Ying}. The final density matrix of the game after the players execute their
moves is given by%
\begin{equation}
\rho _{f}^{\prime }=\frac{\left( U_{AB}\otimes U_{C}\right) \rho _{f}\left(
U_{AB}\otimes U_{C}\right) ^{\dagger }}{\text{\textrm{Tr}}(\left(
U_{AB}\otimes U_{C}\right) \rho _{f}\left( U_{AB}\otimes U_{C}\right)
^{\dagger })},  \label{4}
\end{equation}%
where the trace operation in the denominator ensures that the output is
normalized and represents the final density matrix of the game. In Eq. \ref%
{4}, $\rho _{f}$ is given by Eq. \ref{1}. The operator $U_{AB}=\sqrt{q}%
I\otimes I+\sqrt{1-q}\sigma _{x}\otimes \sigma _{x}$, represents the
strategy of the two cooperators and $U_{C}=\sqrt{r}I+\sqrt{1-r}\sigma _{x}$,
is the strategy of the third player. The payoff functions of the players are
given by \cite{Iqbal2}%
\begin{equation}
P_{A,B,C}(p,q,r)=\mathrm{Tr}(P_{A,B,C}^{\mathrm{oper}}\rho _{f}^{\prime }),
\label{5}
\end{equation}%
where $P_{A,B,C}^{\mathrm{{\mathrm{oper}}}}$ are the payoff operators for
players $A$, $B$ or $C$, which are given by%
\begin{equation}
P_{A,B,C}^{\mathrm{{\mathrm{oper}}}}=\sum\limits_{i=1}^{8}\left( \alpha
_{i},\beta _{i},\gamma _{i}\right) \times \rho _{ii}^{\prime },  \label{6}
\end{equation}%
with $\rho _{ii}^{\prime }$ are the diagonal elements of the final density
matrix $\rho _{f}^{\prime }$ of the game. $\alpha _{i}$'s $\beta _{i}$'s and
$\gamma _{i}$'s are the elements of the payoff matrix of the three players
game. In Eq. \ref{6}, $\alpha _{i}$'s correspond to the payoff operator $%
P_{A}^{\mathrm{{\mathrm{oper}}}}$of player $A$, $\beta _{i}$'s correspond to
the payoff operator $P_{B}^{\mathrm{{\mathrm{oper}}}}$ of player $B$ and $%
\gamma _{i}$'s correspond to the payoff operator $P_{C}^{\mathrm{{\mathrm{%
oper}}}}$ of player $C$ respectively. According to the rules of the game,
the values of the matrix elements $\alpha _{i}$'s of player $A$ become%
\begin{eqnarray}
\alpha _{1} &=&\alpha _{8}=0,  \nonumber \\
\alpha _{2} &=&\alpha _{3}=\alpha _{6}=\alpha _{7}=1,  \nonumber \\
\alpha _{4} &=&\alpha _{5}=-2.  \label{7}
\end{eqnarray}%
Similarly, the values of $\beta _{i}$'s and $\gamma _{i}$'s for players $B$
and $C$ are, respectively, given as
\begin{eqnarray}
\beta _{1} &=&\beta _{8}=0,  \nonumber \\
\beta _{2} &=&\beta _{4}=\beta _{5}=\beta _{7}=1,  \nonumber \\
\beta _{3} &=&\beta _{6}=-2,
\end{eqnarray}%
\begin{eqnarray}
\gamma _{1} &=&\gamma _{8}=0,  \nonumber \\
\gamma _{3} &=&\gamma _{4}=\gamma _{5}=\gamma _{6}=1,  \nonumber \\
\gamma _{2} &=&\gamma _{7}=-2.
\end{eqnarray}

\subsection{Results and discussion for three players game}

In this section, we present and discuss the results of our calculation
obtained under the action of amplitude damping, depolarizing and phase
damping channels for the three players game. In case of amplitude damping
channel, the payoff function of cooperators $A$ and $B$ is obtained as%
\begin{eqnarray}
P_{A,B}^{\mathrm{AD}} &=&\frac{%
\begin{array}{c}
(-q-r+2qr)\left[ 1-2p\left( 1-p\right) \left( 1-\cos \theta \right) \right]
-2(1-p) \\
\times \sqrt{qr(1-p)(1-q)(1-r)}\sin \theta%
\end{array}%
}{-1+4(p-1)\sqrt{qr(1-p)(1-q)(1-r)}\sin \theta }.  \nonumber \\
&&  \label{8}
\end{eqnarray}%
Maximizing $P_{A,B}^{\mathrm{AD}}$ with respect to $q$, and $r$, we get $q=r=%
\frac{1}{2}$. This result is independent both from entanglement parameter $%
\theta $ and decoherence parameter $p$. Using these values of $q$ and $r$ in
Eq. \ref{8}, the maximum payoff of cooperators becomes%
\begin{equation}
P_{A,B,\max }^{\mathrm{AD}}=\frac{\left[ 1-2p\left( 1-p\right) \left( 1-\cos
\theta \right) +\left( 1-p\right) ^{3/2}\sin \theta \right] }{2(1+\left(
1-p\right) ^{3/2}\sin \theta )}.  \label{9}
\end{equation}%
\begin{figure}[h]
\begin{center}
\includegraphics[scale=0.6]{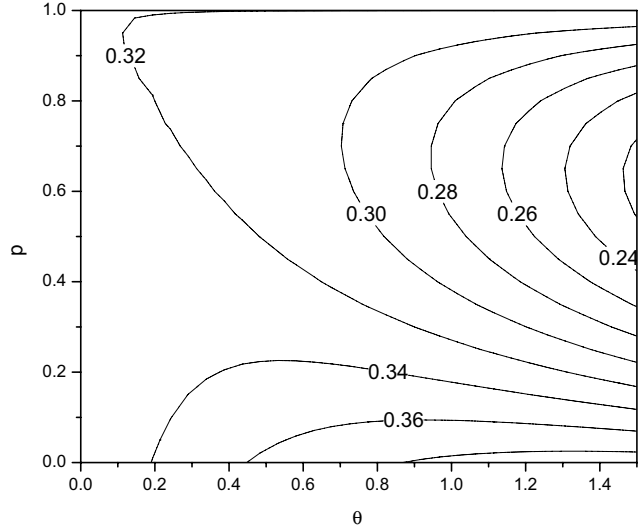}
\end{center}
\caption{The payoff of a player $A$($B$) is plotted against the decoherence parameter $p$ and entanglement parameter $\theta$ for amplitude damping channel with $q=r=0.2$.}\label{Figure1}%
\end{figure}
Unlike the equilibrium payoff in the
classical form of the three players game, this payoff depends both on
entanglement parameter $\theta $ and decoherence parameter $p$. In figure $1$%
, the dependence of cooperators' payoff on both entanglement and decoherence
parameters is shown in the form of a density plot. It is seen that for a
maximally entangled initial state of the game, the payoffs of cooperators
are minimum, when the decoherence parameter has values in the range from $%
0.45$ to $0.72$. Whereas for unentangled initial state, the presence of
decoherence parameter does not affect the payoff considerably for the entire
range of its values. For other values of entanglement parameter, the
presence of decoherence damps the payoff as compared to undecohered case.

The payoff of player $C$ is obtained as%
\begin{eqnarray}
P_{C}^{\mathrm{AD}} &=&-\frac{%
\begin{array}{c}
2\left[ \left( 2qr-q-r\right) \left( -1+2p\left( 1-p\right) \left( 1-\cos
\theta \right) \right) \right] +4\left( 1-p\right) \\
\times \sqrt{qr\left( 1-p\right) \left( 1-q\right) \left( 1-r\right) }\sin
\theta%
\end{array}%
}{+1+4\left( 1-p\right) \sqrt{qr\left( 1-p\right) \left( 1-q\right) \left(
1-r\right) }\sin \theta }.  \nonumber \\
&&  \label{10}
\end{eqnarray}%
Maximizing player $C$'s payoff with respect to $q$ and $r,$ and using their
values in Eq. \ref{10}, the equilibrium payoff becomes%
\begin{equation}
P_{C,\max }^{\mathrm{AD}}=-\frac{1-2p(1-p)\left( 1-\cos \theta \right)
+(1-p)^{3/2}\sin \theta }{1+(1-p)^{3/2}\sin \theta }.  \label{11}
\end{equation}

In case of depolarizing channel, the payoff function of the cooperators
becomes%
\begin{equation}
P_{A,B}^{\mathrm{DP}}=\frac{(3-4p)^{2}(3q+3r-6qr+2(3-4p)\sqrt{qr(1-q)(1-r)}%
\sin \theta )}{27-4(-3+4p)^{3}\sqrt{qr(1-q)(1-r)}\sin \theta }.
\end{equation}

The maximization of $P_{A,B}^{\mathrm{DP}}$ with respect to $q$ and $r$,
leads to $q=r=\frac{1}{2}$, and the maximum payoff of cooperators becomes

\begin{equation}
P_{A,B,\max }^{\mathrm{DP}}=\frac{(3-4p)^{2}(3+(3-4p)\sin \theta )}{%
54-2(-3+4p)^{3}\sin \theta }.  \label{13}
\end{equation}%
\begin{figure}[h]
\begin{center}
\includegraphics[scale=0.6]{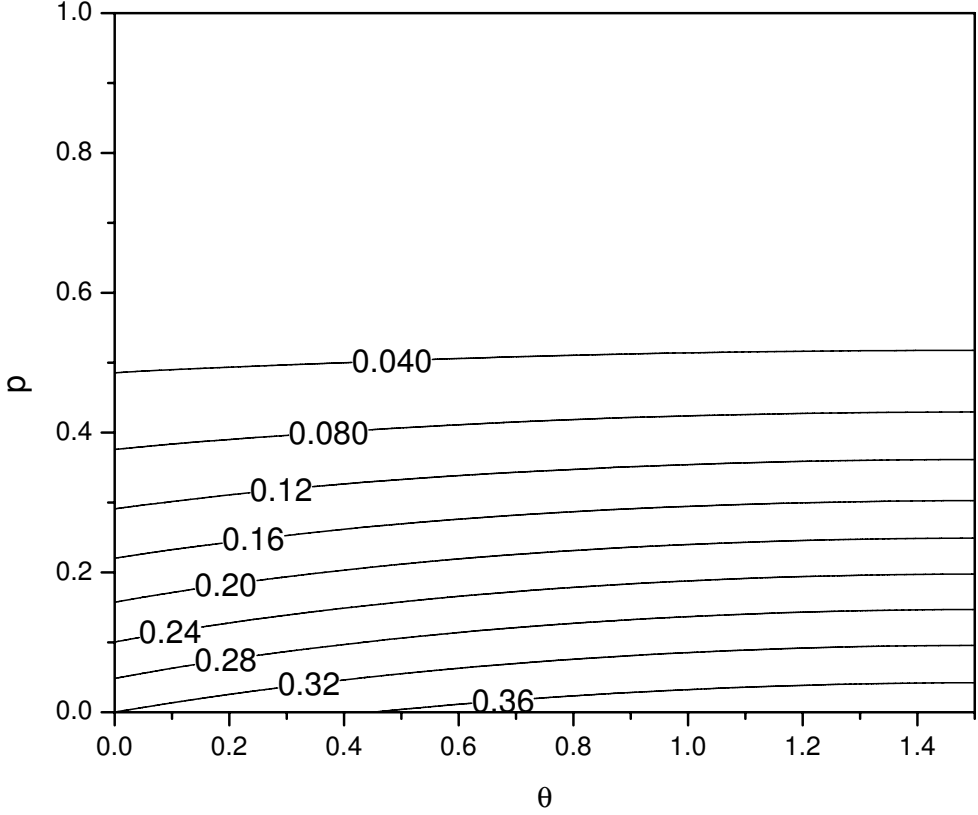}
\end{center}
\caption{The payoff of player $A$($B$) is plotted against the decoherence parameter $p$ and entanglement parameter $\theta $ for depolarizing channel with $q=r=0.2$.}\label{Figure2}%
\end{figure}
The dependence of the payoff on
decoherence and entanglement parameters shows that the behavior of the game
is different both from undecohered and unentangled initial state cases. In
this case, the payoff of cooperators against decoherence and entanglement
parameters is illustrated in figure $2$ as a density plot. In the range of
large values of decoherence parameter, the payoffs of the players vanish
irrespective of the degree of entanglement in the initial state of the game.
Thus for a fully decohered depolarizing channel the advantage of
entanglement, contrary to small values of decoherence parameter, in the
initial state of the game vanishes.

The payoff of player $C$ is given by%
\begin{equation}
P_{C}^{\mathrm{DP}}=-\frac{2(3-4p)^{2}(3q+3r-6qr+2(3-4p)\sqrt{qr(1-q)(1-r)}%
\sin \theta )}{27-4(-3+4p)^{3}\sqrt{qr(1-q)(1-r)}\sin \theta )}.  \label{14a}
\end{equation}

The maximized payoff of player $C$ becomes%
\begin{equation}
P_{C,\max }^{\mathrm{DP}}=-\frac{(3-4p)^{2}(3+(3-4p)\sin \theta }{%
27-(-3+4p)^{3}\sin \theta }.
\end{equation}%
The payoff of cooperators under the action of phase damping channel is given
by%
\begin{equation}
P_{A,B}^{\mathrm{PD}}=\frac{q+r-2qr+2(1-p)\sqrt{qr(1-p)(1-q)(1-r)}\sin
\theta }{1+4\left( 1-p\right) \sqrt{qr(1-p)(1-q)(1-r)}\sin \theta }.
\label{16}
\end{equation}%
The maximized payoff of the cooperators happens at $q=r=1/2$ and is given by%
\begin{equation}
P_{A,B,\max }^{\mathrm{PD}}=\frac{1}{2}.  \label{17}
\end{equation}%
\begin{figure}[h]
\begin{center}
\includegraphics[scale=0.6]{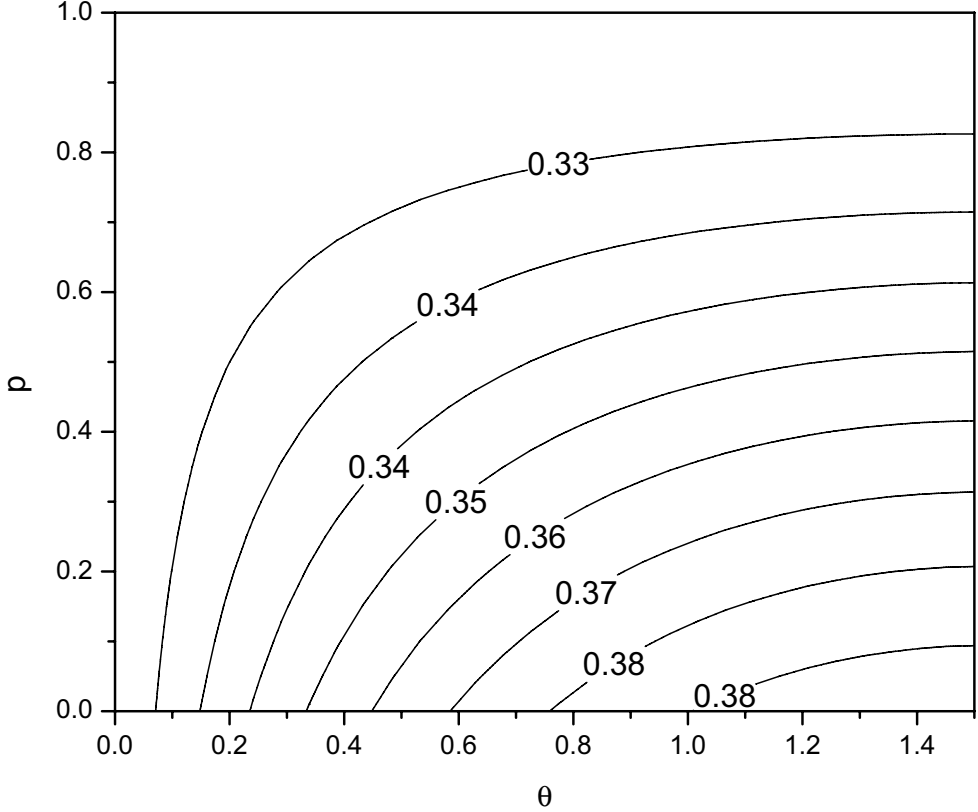}
\end{center}
\caption{The payoff of player $A$($B$) is plotted against the decoherence parameter $p$ and entanglement parameter $\theta$ for phase damping channel with $q=r=0.2$.}\label{Figure3}%
\end{figure}
In figure $3$, we plot the payoff of
cooperators as a function of decoherence and entanglement parameters for
phase damping channel. It can be seen that in the absence of entanglement in
the initial state, the payoff is minimum for the entire range of decoherence
parameter. Similarly, for highly decohered channel, the degree of
entanglement does not effect the outcome of the game and the payoff of
cooperators in this range of decoherence parameter remains minimum.

The payoff of player $C$ is negative and twice the payoff function of a
cooperator, that is,%
\begin{equation}
P_{C}^{\mathrm{PD}}=-2P_{A,B}^{\mathrm{PD}}.  \label{18}
\end{equation}%
\begin{figure}[h]
\begin{center}
\includegraphics[scale=1.2]{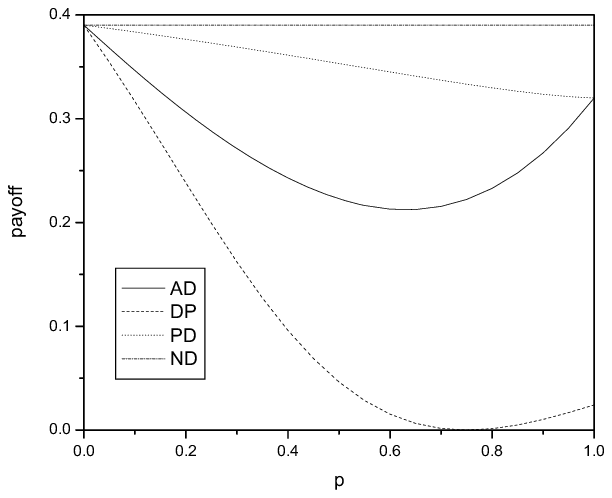}
\end{center}
\caption{The payoff of player $A$($B$) for all the three channels is plotted against the decoherence parameter $p$ when the initial state is maximally entangled with $q=r=0.2$. The labels $\mathrm{%
AD}$, $\mathrm{DP}$, $\mathrm{PD}$ and $\mathrm{ND}$ stand for amplitude
damping, depolarizing, phase damping and no damping cases respectively.}\label{Figure4}%
\end{figure}
The superscripts $\mathrm{AD}$, $%
\mathrm{DP}$ and $\mathrm{PD}$ in the above relations stand for amplitude
damping, depolarizing and phase damping channels respectively. As the sum of
payoffs of the players under all the three channels is zero, the game in its
quantum form with decoherence is a zero sum game. In the classical form of
the game, the maximum values of payoffs that define the Nash equilibrium of
the game is a fixed point. Whereas in the presence of decoherence, the Nash
equilibrium under the action of amplitude and depolarizing channels is a
function of both decoherence parameter $p$ and entanglement parameter $%
\theta $. The effect of decoherence on the payoff of player $A$ ($B$) for
the maximally entangled initial state for all the three channels is shown in
figure $4$. It is seen that for a highly decohered case, the amplitude
damping and phase damping channels reduce the outcome of the game to the
same value. However, heavy damping is observed in the case of amplitude
damping channel as compare to damping in the case of phase damping channel
for $p$ lesser than $1$. The depolarizing channel ends the game with no
payoffs around a $75\%$ decoherence. In figure $5$, we plot the payoff of
cooperators with and without decoherence against entanglement angle for a $%
50\%$ decoherence. It is seen that the channels damp the payoff for the
entire range of entanglement parameter in comparison to undecohered case.
However, the phase damping channel makes better off the cooperators than the
other two channels in the range of large values of entanglement parameter.
The amplitude damping channel results in high degradation in the range of
large values of entanglement parameter. It can also be seen that under the
influence of depolarizing channel, the decoherence results in heavy damping
of the payoff. Furthermore, the effect of entanglement on the payoff
function for depolarizing channel almost vanishes. It can also be shown that
the game becomes a no-loss no-gain game for the entire range of entanglement
parameter when the channel is highly decohered.
\begin{figure}[h]
\begin{center}
\includegraphics[scale=1.2]{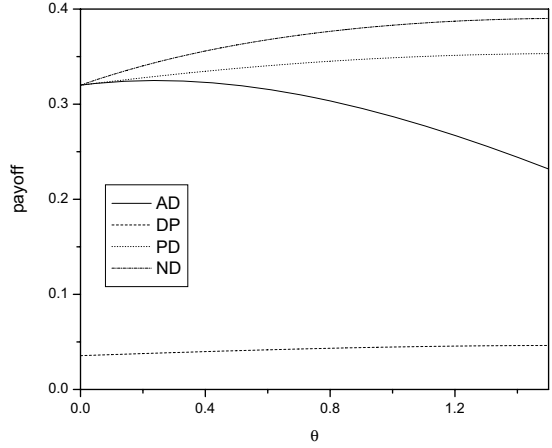}
\end{center}
\caption{The payoff of player $A$($B$) for all the three channels
is plotted against the entanglement parameter $\protect\theta $ for
decoherence parameter $p=0.5$ with $q=r=0.2$. The labels $\mathrm{AD}$, $%
\mathrm{DP}$, $\mathrm{PD}$ and $\mathrm{ND}$ stand for amplitude damping,
depolarizing, phase damping and no damping cases respectively.}\label{Figure5}%
\end{figure}

\section{Decoherence in four players cooperative game}

In this section, we study the effect of decoherence on four players
cooperative game, using quantum superposed operator method, by using the
three quantum channels as in the case of three players cooperative game. The
game space in this case is a sixteen dimensional Hilbert space. We consider
the initial state of the game to be $|\psi \rangle =\cos \theta
/2|0000\rangle +\sin \theta /2|1111\rangle $. The strategy of the two
cooperating players is $U_{AB}=\sqrt{q}I\otimes I+\sqrt{1-q}\sigma
_{x}\otimes \sigma _{x}$, whereas for the other two players the strategies
are respectively given by $U_{C}=\sqrt{r}I+\sqrt{1-r}\sigma _{x}$ and $U_{D}=%
\sqrt{s}I+\sqrt{1-s}\sigma _{x}$. The final density matrix of the game after
the players execute their strategies is given by%
\begin{equation}
\rho _{f}^{\prime }=\frac{\left( U_{AB}\otimes U_{C}\otimes U_{D}\right)
\rho _{f}\left( U_{AB}\otimes U_{C}\otimes U_{D}\right) ^{\dagger }}{\text{%
\textrm{Tr}}(\left( U_{AB}\otimes U_{C}\otimes U_{D}\right) \rho _{f}\left(
U_{AB}\otimes U_{C}\otimes U_{D}\right) ^{\dagger })},  \label{19}
\end{equation}%
where $\rho _{f}$ is given by Eq. \ref{1}. The payoff functions of the
players are given by Eq. \ref{5} with payoff operator given by%
\begin{equation}
P_{A,B,C,D}^{\mathrm{{\mathrm{oper}}}}=\sum\limits_{i=1}^{16}\left( \alpha
_{i},\beta _{i},\gamma _{i},\delta _{i}\right) \times \rho _{ii}^{\prime }.
\label{20}
\end{equation}%
The payoff operators $P_{A}^{\mathrm{oper}}$, $P_{B}^{\mathrm{oper}}$, $%
P_{C}^{\mathrm{oper}}$ and $P_{D}^{\mathrm{oper}}$ correspond to the matrix
elements $\alpha _{i}$'s, $\beta _{i}$'s, $\gamma _{i}$'s\ and $\delta _{i}$%
's respectively. According to the rules of the game, the matrix elements $%
\alpha _{i}$'s of player $A$ become%
\begin{eqnarray}
\alpha _{1} &=&\alpha _{4}=\alpha _{6}=\alpha _{7}=\alpha _{10}=\alpha
_{11}=\alpha _{13}=\alpha _{16}=0,  \nonumber \\
\alpha _{2} &=&\alpha _{3}=\alpha _{5}=\alpha _{12}=\alpha _{14}=\alpha
_{15}=1,  \nonumber \\
\alpha _{8} &=&\alpha _{9}=-3.  \label{21}
\end{eqnarray}%
For the other three players, the matrix elements $\beta _{i}$'s, $\gamma
_{i} $'s\ and $\delta _{i}$'s are given by%
\begin{eqnarray}
\beta _{1} &=&\beta _{4}=\beta _{6}=\beta _{7}=\beta _{10}=\beta _{11}=\beta
_{13}=\beta _{16}=0,  \nonumber \\
\beta _{2} &=&\beta _{3}=\beta _{8}=\beta _{9}=\beta _{14}=\beta _{15}=1,
\nonumber \\
\beta _{5} &=&\beta _{12}=-3,
\end{eqnarray}%
\begin{eqnarray}
\gamma _{1} &=&\gamma _{4}=\gamma _{6}=\gamma _{7}=\gamma _{10}=\gamma
_{11}=\gamma _{13}=\gamma _{16}=0,  \nonumber \\
\gamma _{2} &=&\gamma _{5}=\gamma _{8}=\gamma _{9}=\gamma _{12}=\gamma
_{15}=1,  \nonumber \\
\gamma _{3} &=&\gamma _{14}=-3,
\end{eqnarray}%
\begin{eqnarray}
\delta _{1} &=&\delta _{4}=\delta _{6}=\delta _{7}=\delta _{10}=\delta
_{11}=\delta _{13}=\delta _{16}=0,  \nonumber \\
\delta _{3} &=&\delta _{5}=\delta _{8}=\delta _{9}=\delta _{12}=\delta
_{14}=1,  \nonumber \\
\delta _{2} &=&\delta _{15}=-3.
\end{eqnarray}%
The payoffs of players for the case of amplitude damping channel become%
\begin{eqnarray}
P_{A,B}^{\mathrm{AD}} &=&\frac{%
\begin{array}{c}
\left( r+s-2rs\right) \left[ 1-2p\left( 1-p\right) (1-\cos \theta )\right]
+4\left( 1-p\right) ^{2} \\
\times \sqrt{qrs\left( 1-q\right) \left( 1-r\right) \left( 1-s\right) }\sin
\theta%
\end{array}%
}{1+8\left( 1-p\right) ^{2}\sqrt{qrs\left( 1-q\right) \left( 1-r\right)
\left( 1-s\right) }\sin \theta },  \nonumber \\
P_{C}^{\mathrm{AD}} &=&\frac{%
\begin{array}{c}
\lbrack s-4qs+r(-3+4q+2s)][(1-2p(1-p)(1-\cos \theta )]-4(1-p)^{2} \\
\times \sqrt{qrs(1-q)(1-r)(1-s)}\sin \theta%
\end{array}%
}{1+8\left( 1-p\right) ^{2}\sqrt{qrs\left( 1-q\right) \left( 1-r\right)
\left( 1-s\right) }\sin \theta },  \nonumber \\
P_{D}^{\mathrm{AD}} &=&\frac{%
\begin{array}{c}
\lbrack r(-1+4q-2s)+(3-4q)s][-1+2p(1-p)(1-\cos \theta )]-4(1-p)^{2} \\
\times \sqrt{qrs(1-q)(1-r)(1-s)}\sin \theta%
\end{array}%
}{1+8\left( 1-p\right) ^{2}\sqrt{qrs\left( 1-q\right) \left( 1-r\right)
\left( 1-s\right) }\sin \theta }.  \nonumber \\
&&  \label{22}
\end{eqnarray}%
The maximization of these payoffs gives $q=r=s=\frac{1}{2}$ and the
corresponding maximum payoffs of the players become%
\begin{eqnarray}
P_{A,B,\max }^{\mathrm{AD}} &=&\frac{1-2p(1-p)(1-\cos \theta )+(1-p)^{2}\sin
\theta }{2[1+(1-p)^{2}\sin \theta ]}  \nonumber \\
P_{C,D,\max }^{\mathrm{AD}} &=&-\frac{1-2p(1-p)(1-\cos \theta
)+(1-p)^{2}\sin \theta }{2[1+(1-p)^{2}\sin \theta ]}  \label{23}
\end{eqnarray}%
The payoffs of players under the action of depolarizing channel are given as%
\begin{eqnarray}
P_{A,B}^{\mathrm{DP}} &=&\frac{(3-4p)^{2}[9(r+s-2rs)+4(3-4p)^{2}\sqrt{%
qrs(1-q)(1-r)(1-s)}\sin \theta ]}{81+8(3-4p)^{4}\sqrt{qrs(1-q)(1-r)(1-s)}%
\sin \theta },  \nonumber \\
P_{C}^{\mathrm{DP}} &=&\frac{%
\begin{array}{c}
(3-4p)^{2}[9(s-4qs-3r+4qr+2rs)-4(3-4p)^{2} \\
\times \sqrt{qrs(1-q)(1-r)(1-s)}\sin \theta ]%
\end{array}%
}{81+8(3-4p)^{4}\sqrt{qrs(1-q)(1-r)(1-s)}\sin \theta },  \nonumber \\
P_{D}^{\mathrm{DP}} &=&-\frac{%
\begin{array}{c}
(3-4p)^{2}[9(-r-4qs+3s+4qr-2rs)+4(3-4p)^{2} \\
\times \sqrt{qrs(1-q)(1-r)(1-s)}\sin \theta ]%
\end{array}%
}{81+8(3-4p)^{4}\sqrt{qrs(1-q)(1-r)(1-s)}\sin \theta }.  \nonumber \\
&&  \label{24}
\end{eqnarray}%
The payoffs of players when the game is played under the action of phase
damping channel can be written as%
\begin{eqnarray}
P_{A,B}^{\mathrm{PD}} &=&\frac{r+s-2rs+4(1-p)^{2}\sqrt{qrs(1-q)(1-r)(1-s)}%
\sin \theta }{1+8(1-p)^{2}\sqrt{qrs(1-q)(1-r)(1-s)}\sin \theta },  \nonumber
\\
P_{C}^{\mathrm{PD}} &=&\frac{s-4qs-3r+4qr+2rs-4(1-p)^{2}\sqrt{%
qrs(1-q)(1-r)(1-s)}\sin \theta }{1+8(1-p)^{2}\sqrt{qrs(1-q)(1-r)(1-s)}\sin
\theta },  \nonumber \\
P_{D}^{\mathrm{PD}} &=&\frac{r-4qr-3s+4qs+2rs-4(1-p)^{2}\sqrt{%
qrs(1-q)(-1+r)(-1+s)}\sin \theta }{1+8(1-p)^{2}\sqrt{qrs(1-q)(1-r)(1-s)}\sin
\theta }.  \nonumber \\
&&  \label{25}
\end{eqnarray}%
A similar behavior of the players' payoffs is seen as in the case of
three-player game under decoherence.

\section{Conclusion}

Cooperative three and four player quantum games influenced by different
noise channels are analyzed. The advantage of quantum entanglement in the
initial state of the game for cooperators is adversely affected. For a given
decoherence level, the cooperators are better off under the action of phase
damping channel in the range of larger values of entanglement angle as
compared to the other two channels. In the case of amplitude damping
channel, for a fixed value of decoherence parameter, a decrease in payoff of
cooperators is observed with the increasing value of entanglement parameter.
The game becomes a no-payoff game around a decoherence of $75\%$
irrespective of the degree of entanglement in the case of depolarizing
channel. For a fully decohered case, the amplitude damping and phase damping
channels reduce the outcome of the game to the same value. Furthermore, for
a maximally entangled initial state under the action of amplitude damping
channel the payoffs of cooperators reaches to a minimum at $p=0.7$ and
increase again till the channel becomes fully decohered. In brief, the
decoherence makes the cooperators' payoffs worse off both in three players
and four players cooperative game.\newline

\textbf{acknowledgement}
One of the authors (Salman Khan) is thankful to World Federation of
Scientist for partial financial support under the National Scholarship
Program for Pakistan.

{\huge Figures Captions}\newline
Figure $1$: The payoff of player $A$($B$) is plotted against the decoherence
parameter $p$ and entanglement parameter $\theta $ for amplitude damping
channel with $q=r=0.2$.\newline
Figure $2$: The payoff of player $A$($B$) is plotted against the decoherence
parameter $p$ and entanglement parameter $\theta $ for depolarizing channel
with $q=r=0.2$.\newline
Figure $3$: The payoff of player $A$($B$) is plotted against the decoherence
parameter $p$ and entanglement parameter $\theta $ for phase damping channel
with $q=r=0.2$.\newline
Figure $4$: The payoff of player $A$($B$) for all the three channels is
plotted against the decoherence parameter $p$ when the initial state is
maximally entangled with $q=r=0.2$. The labels $\mathrm{AD}$, $\mathrm{DP}$,
$\mathrm{PD}$ and $\mathrm{ND}$ stand for amplitude damping, depolarizing,
phase damping and no damping cases respectively.\newline
Figure $5$: The payoff of player $A$($B$) for all the three channels is
plotted against the entanglement parameter $\theta $ for decoherence
parameter $p=0.5$ with $q=r=0.2$. The labels $\mathrm{AD}$, $\mathrm{DP}$, $%
\mathrm{PD}$ and $\mathrm{ND}$ stand for amplitude damping, depolarizing,
phase damping and no damping cases respectively.\newline
{\huge Table Caption}\newline
Table $1$. A single qubit Kraus operators for amplitude damping channel,
phase

damping channel and depolarizing channel.


\begin{thebibliography}{99}
\bibitem{Von} Von Neumann, J.: Appl. Math. Ser. \textbf{12} 36 (1951)

\bibitem{Broom1} Broom, M., Cannings, C. Vicker, G. T.: Bull. Math. Biol.
\textbf{59,} 931 (1997)

\bibitem{Broom2} Broom M.: Comments Theor. Biol. \textbf{8}, 103 (2003)

\bibitem{Hofbauer} Hofbauer J., Sigmund K.: Evolutionary Games and
Population Dynamics, Cambridge Univ. Press, Cambridge (1998)

\bibitem{Piotrowski} Piotrowski, E. W., Sladkowski, J.: Physica A \textbf{312%
}, 208 (2002)

\bibitem{Baaquie} Baaquie, B. E.: Phy. Rev. E \textbf{64}, 056122 (2002)

\bibitem{Eisert} Eisert, J., Wilkens, M., Lewenstein, M.: Phys. Rev. Lett
\textbf{83}, 3077 (1999)

\bibitem{Benjamin} Benjamin, S. C., Hayden, P. M.: Phys. Rev. Lett. \textbf{%
87}, 069801 (2001)

\bibitem{Marinatto} Marinatto, L., Weber, T.: Phys. Lett. A, \textbf{272},
291 (2000)

\bibitem{Flitney} Flitney, A. P., Abbott, D.: Phys. Rev. A \textbf{65},
062318 (2002)

\bibitem{Li} Li, H., Du, J., Massar, S.: Phys. Lett. A \textbf{306}, 73
(2002)

\bibitem{Lo} Lo, C. F., Kiang, D.: Phys. Lett. A \textbf{318}, 333 (2003)

\bibitem{LiCF} Li, C. F., Zhang, Y. S., Huang, Y. F, Guo, G. C.: Phys. Lett.
A \textbf{280}, 257 (2001)

\bibitem{Johnson} Johnson, N. F.: Phys. Rev. A \textbf{63}, 020302 (2001)

\bibitem{Shuai} Shuai, C. Mao-Fa, F., Xiao-Juan, Z., Xin-Wen, W.,and Ze-Hua,
L.: Commun. Theor. Phys. \textbf{49,} 100 ( 2008).

\bibitem{Hai} Hai-Jun, Z., and Xi-Ming, F.,: Commun. Theor. Phys. \textbf{45}%
, 75 (2006).

\bibitem{Meyer} Meyer, D. A.: Phys. Rev. Lett. \textbf{82}, 1052 (1999)

\bibitem{Zurek} Zurek, W. H., et al.: Phys. Today \textbf{44}, 36 (1991)

\bibitem{Steane} Steane, A. M.: Phys. Rev. Lett. \textbf{77}, 793 (1996)

\bibitem{Deutsch} Deutsch, D., Ekert, A., Josza, R., Macchiavello, C.,
Popescu, S., Sanpera, A.: Phys. Rev. Lett. \textbf{77}, 2818 (1996)

\bibitem{Flitney3} Flitney, A.P., Abbott, D.: J. Phys. A : Math. Gen.
\textbf{38}, 449 (2005)

\bibitem{Salman} Salman Khan., Ramzan, M., Khan, M. K.: Int J. Theor. Phys.
\textbf{49}, 31 (2010)

\bibitem{Salman1} Salman Khan., Ramzan, M., Khan, M. K.:Commun. Theor. Phys.
\textbf{54}, 47 (2010)

\bibitem{Gawron} Gawron, P., Miszczak, J.A., Sladkowski, J.: Int. J. Quant.
Inf. \textbf{6}, 667 (2008)

\bibitem{Gawron2} Gawron, P.: Fluctuation and Noise Letters. 9, 9 (2010)

\bibitem{Chen} Chen, L.K., Ang, H., Kiang, D., Kwek, L.C., Lo, C.F.: Phys.
Lett. A \textbf{316}, 317 (2003)

\bibitem{Xia} Zhu, X., Kuang, L. M.: J. Phys. A: Math. Theor. \textbf{40},
7729 (2007)

\bibitem{Ramzan1} Ramzan, M., Nawaz, A., Toor, A. H., Khan, M. K.: J. Phys.
A: Math. Theor. \textbf{41,} 055307 (2008)

\bibitem{Benjamin2} Benjamin, S.C., Hayden, P.M.: Phys. Rev. A \textbf{64,}
030301 (2001)

\bibitem{Kay} Kay, R., Johnson, N.F., Benjamin S.C.: J. Phys. A: Math. Gen
\textbf{34,} L547 (2001)

\bibitem{Du} Du, J., Li, H., Xu, X., Zhou, X., Han, R.: Phys. Lett. A
\textbf{302,} 229 (2002)

\bibitem{Han} Han, Y, J., Zhang, Y. S., Guo, G. C.: Phys. Lett. A \textbf{295%
}, 61 (2002)

\bibitem{Iqbal} Iqbal, A., Toor, A.H., Phys. Lett. A \textbf{294}/5--6 261
(2002)

\bibitem{Abbott} Flitney, A.P., Abbott, D.: J. Opt. B: Quantum Semiclass.
Opt. \textbf{6}, S860 (2004)

\bibitem{Wang} Chen, Q., Wang, Y., Liu, J.T., Wang, K.L.: Phys. Lett. A
\textbf{327,} 98 (2004)

\bibitem{Hollenberg} Flitney, A.P.,Hollenberg, L.C.L.: Quantum Inf. Comput.
\textbf{7,} 111 (2007)

\bibitem{Iqbal2} Iqbal, A., Toor, A.H.: Phys. Lett. A \textbf{293}, 103
(2002)

\bibitem{Ying} Ying Jun Ma et al.: Phys. Lett. A \textbf{301,} 117 (2002)

\bibitem{Nielson} Nielson M A. Chuang I. L.: Quantum Computation and Quantum
Information, Cambridge University Press, Cambridge (2000)\pagebreak
\end{thebibliography}
\end{document}